\documentclass[prl,twocolumn,showpacs,aps]{revtex4}
\usepackage{graphicx,graphics,color,epsfig}
\usepackage{bm}
\usepackage{amsmath}
\usepackage{amssymb}
\usepackage{epstopdf}

\begin{document}

\title{Vortex states in hole-doped iron-pnictide superconductors}

\author{Yi Gao$^{1}$, Huai-Xiang Huang$^{1,2}$, Chun Chen$^{3}$, C. S. Ting$^{1,3}$, and Wu-Pei Su$^{1}$}
\affiliation{$^{1}$Department of Physics and Texas Center for
Superconductivity, University of Houston, Houston, Texas, 77204,
USA\\
$^{2}$Department of Physics, Shanghai University, Shanghai, 200444,
China\\
$^{3}$Department of Physics, Fudan University, Shanghai,
200433, China}

\begin{abstract}

Based on a phenomenological model with competing spin-density-wave
(SDW) and extended $s-$wave superconductivity, the vortex states in
Ba$_{1-x}$K$_{x}$Fe$_{2}$As$_{2}$ are investigated by solving
Bogoliubov-de Gennes equations. Our result for the optimally doped
compound without induced SDW is in qualitative agreement with recent
scanning tunneling microscopy experiment. We also propose that the
main effect of the SDW on the vortex states is to reduce the
intensity of the in-gap peak in the local density of states and
transfer the spectral weight to form additional peaks outside the
gap.

\end{abstract}


\maketitle

The recent discovery of the iron-pnictide superconductors \cite{I1},
whose parent compounds exhibit long-range antiferromagnetic (AF) or
spin-density-wave (SDW) order similar to the cuprates \cite{I2},
provides another promising group of materials for studying the
interplay between magnetism and superconductivity (SC). Especially,
the hole-doped pnictide superconductors, like
Ba$_{1-x}$K$_{x}$Fe$_{2}$As$_{2}$ \cite{I3}, have emerged as one of
the most important systems due to the availability of large
homogeneous single crystals. The phase diagram \cite{I4} for these
materials indicates that the parent compound, upon cooling through
T$_{N}\sim$ 140K \cite{I3,I7}, develops a static SDW order. By
increasing the doping of potassium, the SDW order is suppressed and
the SC order emerges as the temperature (T) falls below T$_{c}$. The
SDW and SC orders coexist in the underdoped samples \cite{I4}. By
further increasing the potassium concentration to the optimally
doped regime, the SDW order disappears. These experimental results
provide compelling evidence for strong competition between the SDW
and SC orders.

Another key issue here is  the superconducting pairing symmetry.
Experimental results on the pairing symmetry remain highly
controversial, leaving the perspectives ranging from nodeless
\cite{I21-10,s1} to nodal gap structure \cite{d1,d2,d3}. Although
evidence for a nodal gap has been accumulated in LaFePO \cite{d2}
and Ba(FeAs$_{1-x}$P$_{x}$)$_{2}$ \cite{d3} systems, in the K- and
Co-doped 122-family of iron pnictides, the experimental data points
to the existence of isotropic gaps, especially in the optimally
doped samples \cite{s1}. Theoretically it was suggested that the
pairing may be established via inter-pocket scattering of electrons
between the hole pockets (around the $\Gamma$ point) and electron
pockets (around the $M$ point), leading to the so-called extended
$s-$wave (ES) pairing symmetry ($\Delta_{\mathbf{k}} \sim \cos
k_{x}+\cos k_{y}$) \cite{I20}.

In this regard, investigating the vortex states in the
iron-pnictides can provide useful information on the interplay
between the SDW and SC, as well as the pairing symmetry. Recent
scanning tunneling microscopy (STM) measurement on
Ba$_{1-x}$K$_{x}$Fe$_{2}$As$_{2}$ \cite{stm} has revealed, for the
first time in the iron-pnictides, the existence of the Andreev bound
states inside the vortex core with a systematic evolution: a single
conductance peak appears at a negatively-biased voltage at the
vortex center, which gradually evolves into two sub-peaks when
moving away from the center, with a dominant spectral weight at
negative bias. This negatively-biased conductance peak has not been
observed in electron-doped Ba(Fe$_{1-x}$Co$_{x}$)$_{2}$As$_{2}$
\cite{stm2} and is beyond current theoretical predictions where a
peak appears at positive bias in a two-orbital model \cite{stm3} and
at zero bias in a five-orbital model \cite{stm}. Therefore, it is
important to develop a sound theory for the vortex states in the
iron-pnictide superconductors.

In this work, we adopt a phenomenological model with competing SDW
and extended $s-$wave superconductivity (ESSC) to study the vortex
states in Ba$_{1-x}$K$_{x}$Fe$_{2}$As$_{2}$ from the local density
of states (LDOS). We show that the evolution of the resonance peak
in the calculated LDOS is in qualitative agreement with STM
experiment. Moreover, the effect of SDW on the vortex states is also
discussed, which we predict to be measurable by future experiments.

We begin with a phenomenologically effective two-orbital model on a
two-dimensional (2D) lattice which considers the asymmetry of the As
atoms above and below the Fe layer \cite{M55}, where the onsite
interactions are solely responsible for the SDW while the
next-nearest-neighbor (NNN) intraorbital attraction causes the ESSC.
The effective mean-field Hamiltonian can be written as \cite{M4}
$$H=-\sum_{ij,\alpha\beta,\sigma}t^{'}_{ij,\alpha\beta}c^{\dag}_{i\alpha\sigma}c_{j\beta\sigma}$$
$$+\sum_{j\beta\sigma}\big{[}-\mu+Un_{j\beta\bar{\sigma}}+(U-2J_{H})n_{j\bar{\beta}\bar{\sigma}}$$
$$+(U-3J_{H})n_{j\bar{\beta}\sigma}\big{]}c^{\dag}_{j\beta\sigma}c_{j\beta\sigma}$$
$$+\sum_{ij,\alpha\beta}(\Delta_{ij,\alpha\beta}c^{\dag}_{i\alpha\uparrow}c^{\dag}_{j\beta\downarrow}+H.c.).\eqno{(1)}$$
Here $i,j$ are the site indices, $\alpha,\beta=1,2$ are the orbital
indices, $\sigma$ represents the spin, $\mu$ is the chemical
potential, and $n_{j\beta\sigma}=\langle
c^{\dag}_{j\beta\sigma}c_{j\beta\sigma}\rangle$ is the electron
density. $U$ and $J_{H}$ are the onsite intraorbital Hubbard
repulsion and Hund coupling, respectively. Here we have the
interorbital Coulomb interaction $U^{\prime}=U-2J_{H}$ according to
symmetry \cite{M5}.
$\Delta_{ij,\alpha\beta}=\delta_{\alpha\beta}\frac{V}{2}(\langle
c_{j\beta\downarrow}c_{i\alpha\uparrow}\rangle-\langle
c_{j\beta\uparrow}c_{i\alpha\downarrow}\rangle)$ is the intraorbital
spin-singlet ES bond order parameter, where $V$ is the NNN
intraorbital attraction. The reason we adopt this model is its
ability \cite{M4} to qualitatively account for the doping evolution
of the Fermi surface and the asymmetry in the SC coherent peaks as
observed by the angle resolved photo-emission spectroscopy \cite{M2}
and STM \cite{M3} experiments on
Ba(Fe$_{1-x}$Co$_{x}$)$_{2}$As$_{2}$. In the presence of a magnetic
field $B$ perpendicular to the plane, the hopping integral can be
expressed as
$t^{'}_{ij,\alpha\beta}=t_{ij,\alpha\beta}$exp$[i\frac{\pi}{\Phi_{0}}\int_{j}^{i}\mathbf{A}(\mathbf{r})\cdot
d\mathbf{r}]$, where $\Phi_{0}=hc/2e$ is the superconducting flux
quantum, and $\mathbf{A}=(-By,0,0)$ is the vector potential in the
Landau gauge. Following Ref. \cite{M55}, we have
\begin{equation}
t_{ij,\alpha\beta}=\begin{cases}
t_{1}&\text{$\alpha=\beta,i=j\pm\hat{x}(\hat{y})$},\\
\frac{1+(-1)^{j}}{2}t_{2}+\frac{1-(-1)^{j}}{2}t_{3}&\text{$\alpha=\beta,i=j\pm(\hat{x}+\hat{y})$},\\
\frac{1+(-1)^{j}}{2}t_{3}+\frac{1-(-1)^{j}}{2}t_{2}&\text{$\alpha=\beta,i=j\pm(\hat{x}-\hat{y})$},\\
t_{4}&\text{$\alpha\neq\beta,i=j\pm(\hat{x}\pm\hat{y})$},\\
0&\text{otherwise}\nonumber.
\end{cases}\eqno{(2)}
\end{equation}

The mean-field Hamiltonian (1) can be diagonalized by solving
self-consistently the Bogoliubov-de Gennes (BdG) equations:
$$H=C^{\dag}MC,$$
$$C^{\dag}=(\cdots,c^{\dag}_{j1\uparrow},c_{j1\downarrow},c^{\dag}_{j2\uparrow},c_{j2\downarrow},\cdots),\eqno{(3)}$$
subject to the self-consistency conditions for the electron density
and the ES bond order parameter:
$n_{j\beta\uparrow}=\sum_{k=1}^{L}|Q_{m-1k}|^{2}f(E_{k})$,
$n_{j\beta\downarrow}=1-\sum_{k=1}^{L}|Q_{mk}|^{2}f(E_{k})$ and
$\Delta_{ij,\beta\beta}=\frac{V}{2}\sum_{k=1}^{L}(Q^{*}_{mk}Q_{nk}+Q^{*}_{n+1k}Q_{m-1k})f(E_{k})$.
Here $L=4N_{x}N_{y}$, with $N_{x}/N_{y}$ being the number of lattice
sites along $\hat{x}/\hat{y}$ direction of the 2D lattice.
$m=4(j_{y}+N_{y}j_{x})+2\beta$, $n=4(i_{y}+N_{y}i_{x})+2\alpha-1$
and $Q$ is a unitary matrix that satisfies
$(Q^{\dag}MQ)_{kp}=\delta_{kp}E_{k}$. Here we used $i=(i_{x},i_{y})$
and $j=(j_{x},j_{y})$, with $i_{x},j_{x}=0,1,\ldots,N_{x}-1$ and
$i_{y},j_{y}=0,1,\ldots,N_{y}-1$. The chemical potential $\mu$ is
determined by the doping concentration $x$ through
$\frac{1}{N_{x}N_{y}}\sum_{j\beta\sigma}n_{j\beta\sigma}=2-\frac{x}{2}$.
The ES order parameter at site $j$ is
$\Delta^{'}_{j\beta}=(\Delta^{'}_{j+\hat{x}+\hat{y}j,\beta\beta}+\Delta^{'}_{j-\hat{x}-\hat{y}j,\beta\beta}
+\Delta^{'}_{j+\hat{x}-\hat{y}j,\beta\beta}+\Delta^{'}_{j-\hat{x}+\hat{y}j,\beta\beta})/4$
where
$\Delta^{'}_{ij,\beta\beta}=\Delta_{ij,\beta\beta}$exp$[i\frac{\pi}{\Phi_{0}}\int_{j}^{(i+j)/2}\mathbf{A}(\mathbf{r})\cdot
d\mathbf{r}]$. The LDOS is given by
$\rho_{i}(\omega)=\sum_{k=1}^{L}\sum_{\alpha}\big{[}|Q_{nk}|^{2}\delta(\omega-E_{k})+
|Q_{n+1k}|^{2}\delta(\omega+E_{k})\big{]}$, the supercell technique
is used to calculate the LDOS.

In our calculation, the magnitudes of the parameters are chosen as
$t_{1-4}=1,0.4,-2,0.04$ \cite{M4}, $U=3.7$, $V=-2$ and $T=10^{-4}$.
Magnetic unit cells are introduced where each unit cell accommodates
two superconducting flux quantum and the linear dimension is
$N_{x}\times N_{y}=48\times24$, which is larger than the coherence
length $\xi$ of the iron-pnictides \cite{stm2}. Throughout the
paper, the length and energy are measured in units of the Fe-Fe
distance $a$ and $t_{1}$, respectively. In the following, we focus
on two doping concentrations $x=0.4$ and $0.3$, corresponding to the
optimally doped and underdoped compounds, respectively.

\begin{figure}
\includegraphics[width=1.0\linewidth]{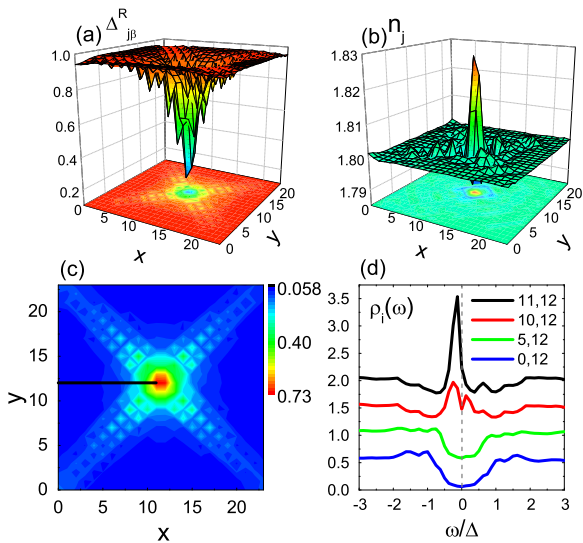}
\caption{\label{figure1} (Color online) Spatial variations of (a)
the reduced ES order parameter $\Delta^{R}_{j\beta}$, (b) electron
density $n_{j}$, and (c) ZE LDOS map plotted on a $24\times24$
lattice. (d) The LDOS at four typical positions along the black cut
in (c): at the vortex center (11,12); within the vortex core while
away from the center (10,12); around the edge of a vortex (5,12);
and far outside a vortex (0,12). The curves in (d) are displaced
vertically for clarity and the gray dashed line indicates the
position of zero energy.}
\end{figure}

At $x=0.4$, first let us choose $J_{H}=0.2U$ such that, at $B=0$,
SDW is completely suppressed and the ES order parameter
$\Delta^{'}_{j\beta}$ is homogeneous in real space. Figures
\ref{figure1}(a) and \ref{figure1}(b) show the spatial variations of
the reduced ES order parameter
$\Delta^{R}_{j\beta}=|\Delta^{'}_{j\beta}/\Delta^{'}_{j\beta}(B=0)|$
and the electron density $n_{j}=\sum_{\beta\sigma}n_{j\beta\sigma}$
plotted on a $24\times24$ lattice. The vortex center is located at
site (11,12) and no SDW is induced. The reduced ES order parameter
$\Delta^{R}_{j\beta}$ vanishes at the vortex center and starts to
increase at the scale of the coherence length $\xi$ to its bulk
value, but the increase is slower along the $\pi/4$ and $3\pi/4$
directions with respect to the underlying lattice. On the other
hand, the electron density $n_{j}$ is strongly enhanced at the
vortex center which is compensated by a depletion of electrons
around two lattice spacings away from the center, after which
$n_{j}$ decays also at the scale of $\xi$ to its bulk value, with no
obviously slow variations along the $\pi/4$ and $3\pi/4$ directions.
The zero-energy(ZE) LDOS plotted in Fig. \ref{figure1}(c) also peaks
at site (11,12) and has the same fourfold rotational symmetry (RS)
as $\Delta^{R}_{j\beta}$. In order to reveal the spatial variation
of LDOS modulated by the vortex, in Fig. \ref{figure1}(d) we plot
the LDOS at four typical positions along the black cut in Fig.
\ref{figure1}(c). As we can see, at the vortex center, there is a
remarkable negative-energy (NE) in-gap peak located at
$-0.125\Delta$, which is precisely the same as observed in Ref.
\cite{stm}. When moving away from the center, the peak will split
into two in-gap peaks with a dominant spectral weight at negative
energy. Finally, the LDOS evolves continuously into its bulk
feature. The in-gap peak and evolution of the LDOS clearly indicate
the existence of the Andreev bound states inside the vortex core,
consistent with Ref. \cite{stm}.

\begin{figure}
\includegraphics[width=1.0\linewidth]{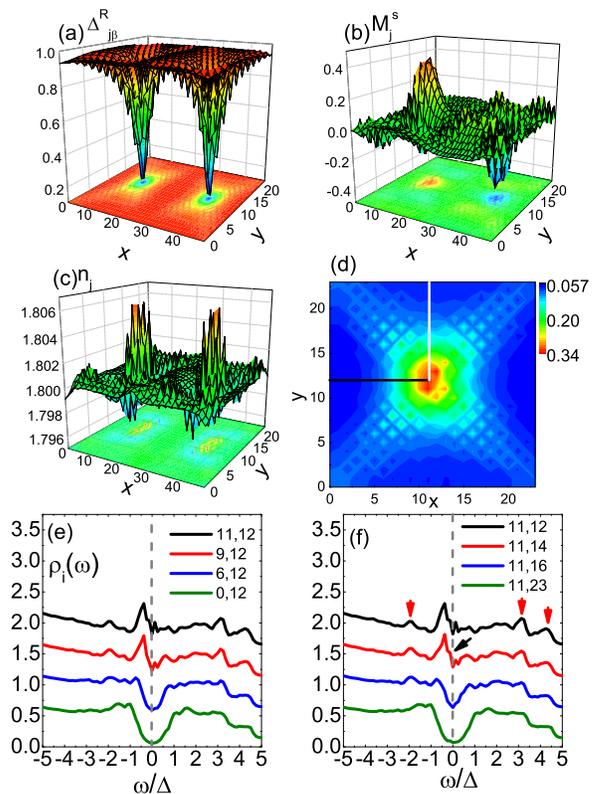}
\caption{\label{figure2} (Color online) Spatial variations of (a)
the reduced ES order parameter $\Delta^{R}_{j\beta}$, (b) staggered
magnetization $M^{s}_{j}$, and (c) electron density $n_{j}$ plotted
on a $48\times24$ lattice. (d) The ZE LDOS map plotted on a
$24\times24$ lattice. (e) The LDOS at four typical positions along
the black cut in (d). (f) is similar to (e), but is plotted along
the white cut.}
\end{figure}

In order to study the effect of induced SDW on the vortex states, we
perform the calculation for $J_{H}=0.23U$. Like the $J_{H}=0.2U$
case, at $B=0$, SDW is completely suppressed and the ES order
parameter $\Delta^{'}_{j\beta}$ is homogeneous in real space, but
the vortex states are fundamentally different from those for
$J_{H}=0.2U$ and they are presented in Fig. \ref{figure2}. As shown
in Fig. \ref{figure2}(a), the vortex center is still located at site
(11,12) where the reduced ES order parameter $\Delta^{R}_{j\beta}$
vanishes, but the size of the vortex core is slightly enlarged
compared to the $J_{H}=0.2U$ case. The induced SDW order parameter
defined as $M^{s}_{j}=(-1)^{j_{y}}(n_{j\uparrow}-n_{j\downarrow})$
displayed in Fig. \ref{figure2}(b) reaches its maximum strength at
the vortex center and decays at the scale of $\xi$ to zero into the
superconducting region. More interestingly, the SDW order parameter
has opposite polarity around two nearest-neighbor vortices along the
$\hat{x}$ direction, thus doubling the period of the translational
symmetry (TS) of the vortex lattice along this direction.
Furthermore, with the induced SDW order, the electron density
$n_{j}$ is only moderately enhanced near the vortex center with a
depletion of electrons around the edge of the vortex core [see Fig.
\ref{figure2}(c)]. The ZE LDOS in Fig. \ref{figure2}(d) also shows a
slightly enlarged vortex core and the RS inside the core is reduced
from fourfold to twofold due to the induced SDW order. Figures
\ref{figure2}(e) and \ref{figure2}(f) are the spatial variations of
the LDOS along the black and white cuts in Fig. \ref{figure2}(d),
respectively. The spectra have only minor differences along the two
cuts, mainly inside the vortex core and close to $\omega=0$
(indicated by the black arrow in Fig. \ref{figure2}(f)). At the
vortex center, besides a NE in-gap peak at $-0.375\Delta$, there is
a small positive-energy in-gap peak at $0.125\Delta$ which does not
exist for $J_{H}=0.2U$. In addition, the intensity of the NE peak is
strongly reduced as compared to the $J_{H}=0.2U$ case and the
spectral weight is transferred to form additional peaks outside the
gap as indicated by the red arrows in Fig. \ref{figure2}(f). When
moving away from the center, the intensities of all these peaks
decrease and finally the LDOS evolves into its bulk feature. By
comparing with the $J_{H}=0.2U$ case, we can identify that those two
in-gap peaks are due to the Andreev bound states while the others
are due to the induced SDW order inside the core.

\begin{figure}
\includegraphics[width=1.0\linewidth]{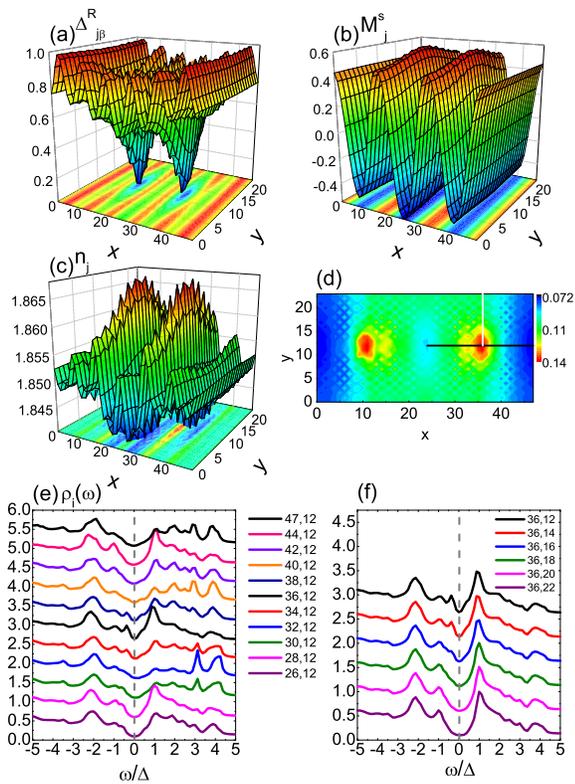}
\caption{\label{figure3} (Color online) (a), (b), (c) and (d) are
similar to Figs. \ref{figure2}(a), \ref{figure2}(b),
\ref{figure2}(c) and \ref{figure2}(d), respectively. (e) The LDOS
along the black cut in (d). (f) is similar to (e), but is plotted
along the white cut.}
\end{figure}

At $x=0.3$, we choose $J_{H}=0.32U$ so that, at $B=0$, the $(\pi,0)$
SDW coexists with the ESSC. The vortex states are plotted in Fig.
\ref{figure3}. The reduced ES order parameter $\Delta^{R}_{j\beta}$
[see Fig. \ref{figure3}(a)] shows a $\hat{y}-$axis oriented
stripe-like feature with a modulation period of $8a$. The size of
the vortex core is further enlarged and elongated along the
$\hat{y}$ direction. Moreover, the two vortex cores are dragged
towards each other along the $\hat{x}$ direction with the vortex
centers located at sites $(15,12)$ and $(32,12)$, thus also doubling
the period of the TS of the vortex lattice along this direction. The
SDW order parameter $M^{s}_{j}$ shown in Fig. \ref{figure3}(b)
behaves like nearly uniform stripes oscillating with a wavelength of
$16a$. The vortex core is pinned at one of the ridges of SDW stripes
where the SDW order is stronger than those at other sites. The
spatial variation of the electron density $n_{j}$ also exhibits a
quasi-one-dimensional charge stripe behavior with a wavelength $8a$,
exactly half that of the SDW along the $\hat{x}$ direction [see Fig.
\ref{figure3}(c)]. The one-dimensional stripe-like modulations in
$\Delta^{R}_{j\beta}$, $M^{s}_{j}$ and $n_{j}$ already exist at
$B=0$, which are quite similar to the cuprates except for a doubling
of the period from $4a$ (for $\Delta^{R}_{j\beta}$ and $n_{j}$) and
$8a$ (for $M^{s}_{j}$) in the cuprates \cite{stripe} to $8a$ and
$16a$ in the iron-pnictides. The origin of such stripes could be
understood in terms of the existence of a nesting wave vector
$q_{A}\sim0.125\pi/a$ connecting the left (right) pieces of the
inner and outer hole pockets around the $\Gamma$ point along the
$k_{x}$ direction. For proper values of $U$, $J_{H}$ and doping,
this wave vector would modulate $M^{s}_{j}$ with SDW stripes along
the $\hat{x}$ direction with period $2\pi/q_{A}=16a$. The ZE LDOS in
Fig. \ref{figure3}(d) also shows an enlarged vortex core, the
doubling of the period of the TS of the vortex lattice along the
$\hat{x}$ direction and the reduced RS from fourfold to twofold.
Interestingly, although the two vortex cores are dragged towards
each other, the ZE LDOS still peaks at sites $(11,12)$ and
$(36,12)$, suggesting that even in the region where
$\Delta^{R}_{j\beta}\neq0$, there are ZE states contributing to the
LDOS. The spatial variations of the LDOS plotted in Figs.
\ref{figure3}(e) and \ref{figure3}(f) show that at the ZE LDOS peak
position, there is a NE in-gap peak at $-0.375\Delta$, whose
intensity is further reduced compared to that in Fig.
\ref{figure2}(f) and the intensity decreases when moving away from
the peak position, indicating that it is due to the Andreev bound
states. There are also additional peaks outside the gap whose
positions are similar to those marked by the red arrows in Fig.
\ref{figure2}(f). Their intensities vary drastically along the black
cut in Fig. \ref{figure3}(d) while they barely change along the
white cut, again suggesting that these peaks are due to the SDW
order.

In summary, we have systematically investigated the vortex states in
Ba$_{1-x}$K$_{x}$Fe$_{2}$As$_{2}$ with the consideration of the
interplay between the SDW and ESSC. In the optimally doped compound
without induced SDW, there is a NE in-gap peak in the LDOS at the
vortex center due to the Adreev bound states, which splits into two
asymmetric in-gap peaks when moving away from the center. The effect
of the induced SDW is mainly to reduce the intensity of the NE
in-gap peak and transfer the spectral weight to form additional
peaks outside the gap. For the underdoped sample where the SDW
coexists with the ESSC, the vortex cores are dragged towards each
other along the $\hat{x}$ direction and the intensity of the NE
in-gap peak is further reduced. The obtained result at $x=0.4$
without induced SDW is in qualitative agreement with experiment and
we propose future experiments on the near optimally doped and
underdoped samples to verify the effect of the SDW on the vortex
states. On the other hand, the disappearance of the Adreev bound
states in electron-doped Ba(Fe$_{1-x}$Co$_{x}$)$_{2}$As$_{2}$ may be
due to the induction of strong SDW order in the vortex states, which
also needs to be verified by future experiments.

{\it Acknowledgments} We thank D. G. Zhang, T. Zhou, C. H. Li, S. H.
Pan and A. Li for helpful discussions. This work was supported by
the Texas Center for Superconductivity and the Robert A. Welch
Foundation under grant numbers E-1070 (Y. Gao and W. P. Su) and
E-1146 (H. X. Huang and C. S. Ting).


\begin{thebibliography}{99}

\bibitem{I1} Y. Kamihara, T. Watanabe, M. Hirano, and H. Hosono, J. Am. Chem. Soc.
\textbf{130}, 3296 (2008).

\bibitem{I2} P. A. Lee, N. Nagaosa, and X. -G. Wen,
Rev. Mod. Phys. \textbf{78}, 17 (2006), and references therein.

\bibitem{I3} M. Rotter, M. Tegel, and D. Johrendt, Phys. Rev. Lett.
\textbf{101}, 107006 (2008).

\bibitem{I4} H. Chen \emph{et al.}, Europhys. Lett.
\textbf{85}, 17006 (2009); R. R. Urbano \emph{et al.},
arXiv:1005.3718 (2010).

\bibitem{I7} G. Wu \emph{et al.}, Europhys. Lett.
\textbf{84}, 27010 (2008); M. Rotter \emph{et al.}, Phys. Rev. B
\textbf{78}, 020503(R) (2008).

\bibitem{I21-10}  C. Liu \emph{et al.},
Phys. Rev. Lett. \textbf{101}, 177005 (2008); K. Nakayama \emph{et
al.}, Europhys. Lett. \textbf{85}, 67002 (2009); D. V. Evtushinsky
\emph{et al.}, Phys. Rev. B \textbf{79}, 054517 (2009); K. Hashimoto
\emph{et al.}, Phys. Rev. Lett. \textbf{102}, 017002 (2009).

\bibitem{s1} H. Ding \emph{et al.}, Europhys. Lett. \textbf{83}, 47001
(2008); X. G. Luo \emph{et al.}, Phys. Rev. B \textbf{80}, 140503(R)
(2009); R. T. Gordon \emph{et al.}, Phys. Rev. Lett. \textbf{102},
127004 (2009).

\bibitem{d1} H.-J. Grafe \emph{et al.}, Phys. Rev. Lett. \textbf{101}, 047003
(2008);

\bibitem{d2} J. D. Fletcher \emph{et al.}, Phys. Rev. Lett. \textbf{102}, 147001 (2009); C. W.
Hicks \emph{et al.}, Phys. Rev. Lett. \textbf{103}, 127003 (2009).

\bibitem{d3} Y. Nakai \emph{et al.}, Phys. Rev. B \textbf{81}, 020503(R)
(2010).

\bibitem{I20} I. I. Mazin \emph{et al.}, Phys. Rev. Lett. \textbf{101}, 057003
(2008); Z.-J. Yao, J.-X. Li, and Z. D. Wang, New J. Phys.
\textbf{11}, 025009 (2009); F. Wang \emph{et al.}, Phys. Rev. Lett.
\textbf{102}, 047005 (2009).

\bibitem{stm} L. Shan \emph{et al.}, arXiv:1005.4038 (2010).

\bibitem{stm2} Y. Yin \emph{et al.}, Phys. Rev. Lett. \textbf{102},
097002 (2009).

\bibitem{stm3} X. Hu, C. S. Ting, and J. X. Zhu, Phys. Rev. B \textbf{80},
014523 (2009); H. M. Jiang, J. X. Li, and Z. D. Wang, Phys. Rev. B
\textbf{80}, 134505 (2009).

\bibitem{M55} Degang Zhang, Phys. Rev. Lett. \textbf{104}, 089702 (2010).

\bibitem{M4} T. Zhou, Degang Zhang, and C. S. Ting, Phys. Rev. B
\textbf{81}, 052506 (2010).

\bibitem{M5} A. M. Ole\'{s} \emph{et al.}, Phys. Rev. B \textbf{72}, 214431 (2005).

\bibitem{M2} Y. Sekiba \emph{et al.}, New J. Phys. \textbf{11}, 025020
(2009).

\bibitem{M3} S. H. Pan \emph{et al.}, private communication.

\bibitem{stripe} Hong-Yi Chen and C. S. Ting, Phys. Rev. B \textbf{68}, 212502 (2003).

\end{thebibliography}
\end{document}